\documentclass[reprint,amsmath,amssymb,aps]{revtex4-2}

\pdfoutput=1

\usepackage[utf8]{inputenc}
\usepackage{graphicx,tikz}
\usepackage{gensymb}
\usepackage{bm}
\usepackage[normalem]{ulem}
\usepackage{hyperref}
\hypersetup{%
    colorlinks=true
}

\graphicspath{{figures/}}

\begin{document}

\preprint{APS/123-QED}

\title{Proximity-Enhanced Magnetocaloric Effect in Ferromagnetic Trilayers}
\author{M.\ Persson}
\email{miltonp@kth.se}
\author{M.\ M.\ Kulyk}
\altaffiliation[Also at ]{Institute of Physics, NASU, 03028 Kyiv, Ukraine.}
\author{A.\ F.\ Kravets}
\altaffiliation[Also at ]{Institute of Magnetism, NASU, 03142 Kyiv, Ukraine.}
\author{V.\ Korenivski}
\affiliation{Nanostructure Physics, Royal Institute of Technology, 10691 Stockholm, Sweden}
\date{\today}

\begin{abstract}
The demagnetization and associated magnetocaloric effect in strong-weak-strong ferromagnetic trilayers, upon a reorientation of the strong ferromagnets from parallel to antiparallel magnetization, is simulated using atomistic spin dynamics. The simulations yield non-trivial spin distributions in the antiparallel state, which in turn allows entropy to be calculated directly. Empirical functional forms are obtained for the magnetization distribution in the spacer, differing significantly from some of the commonly used models. Finally, we find that the magnetocaloric effect in the system can be significantly improved by allowing the local exchange to vary through the spacer, which in practice can be implemented by spatially tailoring the spacer's magnetic dilution.
\end{abstract}

\maketitle

\section{Introduction}
The magnetocaloric effect (MCE) relates to the energy exchange between the phonon and magnon subsystems in a material, manifesting as a decrease in temperature upon demagnetization and, correspondingly, an increase in temperature upon magnetization. The effect was first observed in Nickel by P.\ Weiss and A.\ Piccard in 1917~\cite{Weiss_1917,Smith_2013} and is studied today for cooling applications where it offers more efficiency~\cite{Zimm_1998,Tishin_2003} and can be more environmentally friendly than today's vapor-compression technology, with greater systems' longevity and without high global warming potential (GWP) refrigerants~\cite{Metz_2005}. Many exotic materials exhibiting giant MCE have been found and studied, but when used in their bulk form the fields required to produce substantial values of MCE are well above 1~T~\cite{Ram_2018}. With cooling of electronics being one of the most interesting application, it is vital to reduce these operating fields, and the search has turned to novel nanostructured materials~\cite{Miller_2014}. These may additionally reduce or eliminate the need for rare-earth metals, which are environmentally unfriendly~\cite{Monfared_2014}. 

In a nanostructured trilayer consisting of one free and one pinned ferromagnetic layers, separated by a weakly ferromagnetic spacer, a field of the order of 100~Oe is sufficient to switch the outer ferromagnetic layers from parallel to antiparallel alignment and thereby reduce the effective exchange field in the spacer, which in turn leads to its demagnetization~\cite{Kravets_2012,Kravets_2014}. Such trilayer systems have been rather intensively studied recently, both in experiments and theoretically~\cite{Kravets_2015,Kravets_2016,Fraerman_2015,Magnus_2016,Vdovichev_2018,Polushkin_2019,Kuznetsov_2020,Fraerman_2021,Kuznetsov_2021}. As regards MCE, first discussed by Fraerman and Shereshevskii~\cite{Fraerman_2015}, these studies have mostly been indirect experimentally (via magnetometry) and phenomenological in terms of modeling~\cite{Fraerman_2015,Fraerman_2021,Kuznetsov_2020,Kuznetsov_2021,Polushkin_2019,Vdovichev_2018}, which introduces model-dependent uncertainties in the conclusions drawn.

In this article, the above trilayer system is simulated using the atomistic spin dynamics in the system and the magnetic entropy (MCE) is computed directly from the resulting atomic spin distribution at various temperatures. The results are in good qualitative agreement with direct MCE measurements on a prototype system~\cite{Kulyk_2022}. In addition and quite informatively, our atomistic simulations reveal the profile of the magnetization inside the spacer, which in the antiparallel state is qualitatively different from the functional forms used in the literature. Studying the temperature dependence of the spin profiles leads to a rather intuitive understanding of what these functional forms should be analytically in order to fit the atomistic simulations well at all temperatures. Finally, possible nanostructure optimizations are investigated by allowing the exchange (in practice via tailoring the concentration of the magnetic material) to vary through the spacer, and a significant improvement is found in terms of the entropy change per unit field for a specific gradient-spacer design.

\section{System Description}
The system under study is a trilayer F/f/F with two strongly ferromagnetic layers, F, surrounding a weakly ferromagnetic spacer, f. One of the strongly ferromagnetic layers is pinned, by intrinsic anisotropy or exchange, while the other is free, so that an external magnetic field can oppose the effect of exchange coupling through the spacer and produce an antiparallel (AP) configuration. The idea is that the magnetization of the spacer should be higher (ideally much higher) in the parallel (P) as compared to the AP configuration, so that a significant change in the magnetization could be produced when switching between the two states (P-to-AP and AP-to-P). Furthermore, while for bulk MCE the external field competes directly with thermal disorder and therefore has to be very large, a small field is sufficient to switch the strongly ferromagnetic layers in F/f/F, which is then effectively amplified by the intrinsic exchange competing with thermal disorder within the spacer.

\begin{figure}[!t]
\centering
\includegraphics[scale=0.45]{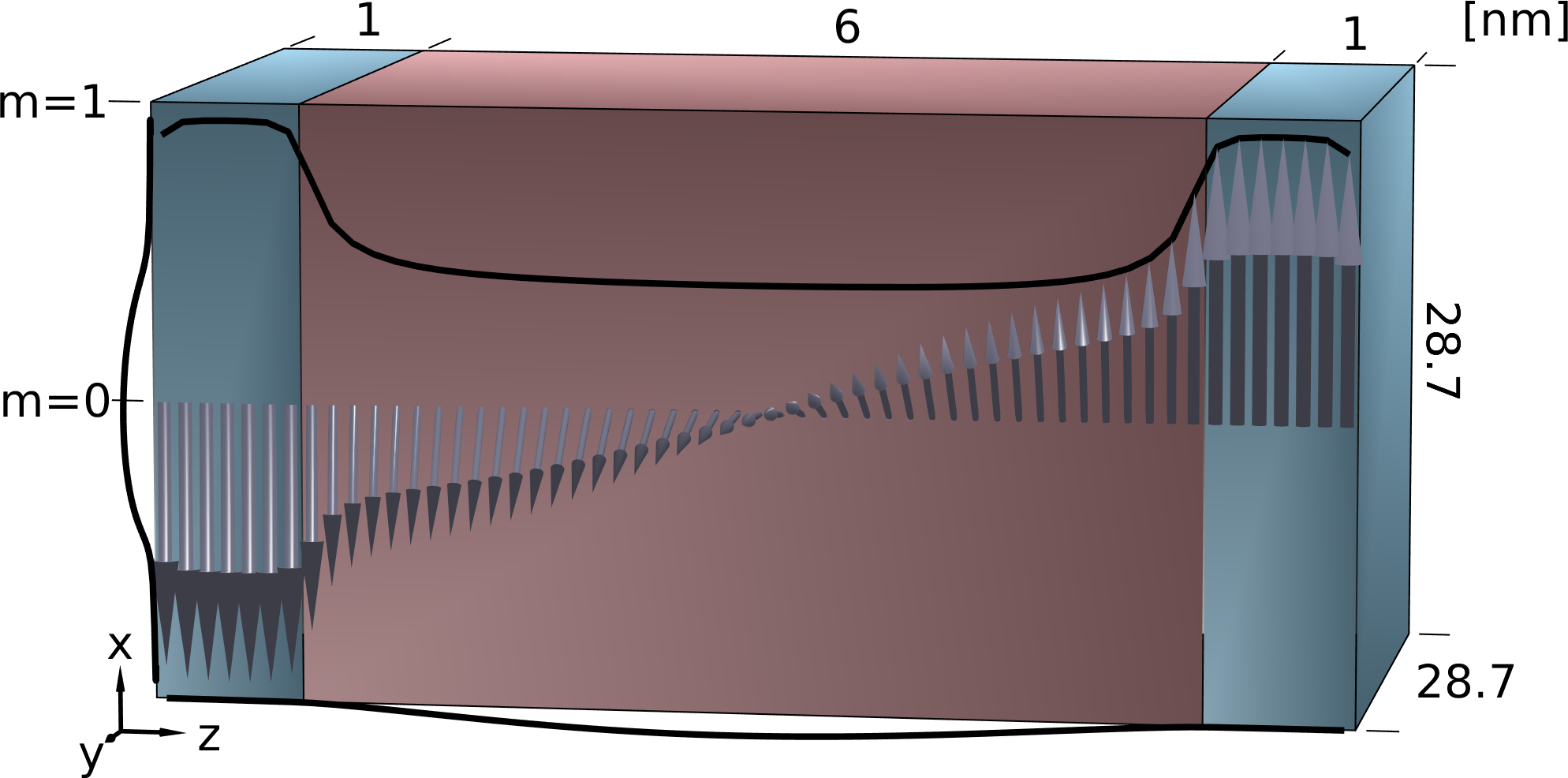}
\caption{\label{fig:sample} Multilayered system under study with two 1~nm thick outer strongly ferromagnetic layers separated by a 6~nm thick weakly ferromagnetic spacer. Arrows show simulated spatial distribution of normalized magnetic moment vectors of each monolayer in the antiparallel F-F magnetization state at a temperature close to $T_{\text{C}}$. Upper black curve shows magnitude of magnetic moments, while similar curves on the left and bottom show projections onto $(x,y)$ and $(y,z)$ planes, respectively.}
\end{figure}

Atomistic spin dynamics simulations are carried out using the {\small VAMPIRE}~\cite{Evans_2014} software package, based on the Landau-Lifshitz-Gilbert equation with Langevin Dynamics. The simulated system is a BCC crystal with $28\times100\times100$ unit cells, periodic exchange in the lateral directions, and the iron lattice parameter of $a=2.87$~\AA, which translates to roughly $8\times29\times29$ nm$^3$ with 1~nm for each of the strong ferromagnetic layers and 6~nm for the weakly ferromagnetic spacer; see Fig.~\ref{fig:sample}.
F.\ Magnus, \textit{et al.}~\cite{Magnus_2016} found that long-range spin-spin interactions need to be taken into account in order to properly reproduce the experimentally observed coupling through thick paramagnetic or weakly ferromagnetic spacers. Here, we test this premise for MCE by comparing the results of our atomistic simulations with the 1st, 2nd, 3rd, and up to 4th nearest neighbors included in the spin-spin exchange, to see if longer range interactions yield a qualitative difference in interpretation. Similar to Ref.~\cite{Magnus_2016} we took the longer-range exchange as decaying algebraically~\cite{Fisher_1972}:
\begin{equation}
    J^{i\text{th nn}} = \left(\frac{r^{\text{1st nn}}}{r^{i\text{th nn}}}\right)^{d+\sigma}J^{\text{1st nn}},
\end{equation}
with $d=3$ and $\sigma = 0.5$. To study the effect of only the range of the spin-spin interactions, the values of the exchange terms were rescaled, so that the total effective exchange field, when all spins are parallel, remains the same. The respective scaling factor, $X_n$, for interactions including up to the $n$th nearest neighbors, satisfies 
\begin{equation}
    J^{\text{1st only}}N^{\text{1st}} = X_n\sum_{i=1}^nJ^{i\text{th}}N^{i\text{th}},
    \label{eqn:rescaling}
\end{equation}
where $N^{i\text{th}}$ is the $i$th coordination number. For the interface exchange between the F and f materials, with the internal exchange values equal to $J_1$ and $J_2$, the harmonic mean was used~\cite{Vansteenkiste_2014}:
\begin{equation}
    J_{1\leftrightarrow2} = 2\frac{J_1J_2}{J_1 + J_2}.
\end{equation}
Deviations from this mean value of the order of 1\% were forced by the rescaling~\eqref{eqn:rescaling}.
Strong easy-plane anisotropy was used for all layers to emulate the demagnetizing fields in the thin film geometry.
The time step used is 1.0~fs which should be adequate for our system with a relatively weak effective exchange in the spacer, even in the vicinity of the Curie temperature, since the resulting spin precession is not exceedingly fast~\cite{Evans_2014}. Furthermore, the spin precession was slowed down and the relaxation sped up by the use of critical damping, $\alpha = 1$.

\section{Results}
\subsection{MCE versus temperature}
In experiments, an external field would be required to switch and compare the magnetization between the parallel and antiparallel states. In our simulations, the outer ferromagnetic layers are instead fixed parallel or antiparallel by giving them easy-axis anisotropy equivalent to 55~T ($5.65\mathrm{E}{-22}$~J/atomic-volume) -- an increase of the  intrinsic iron anisotropy by three orders of magnitude~\cite{Evans_2014}, such that the P and AP states are well defined. The first nearest neighbor exchange was set to $J^{1\text{st}} = 7.05\mathrm{E}{-21}$~J/link in the outer ferromagnetic layers for iron~\cite{Evans_2014}, and the spacer was given 15\% of this value to match the experimentally known $T_{\text{C}}$ for Fe$_{30}$Cr$_{70}$ spacers~\cite{Ravi_Kumar_2015,Ravi_Kumar_2018,Kulyk_2022}. The atomic moments were then set to the iron value of $2.22\mu_{\text{B}}$ in the outer layers and 30\% of this, $0.666\mu_{\text{B}}$, in the spacer.

\begin{figure}[!t]
\centering
\includegraphics[scale=0.45]{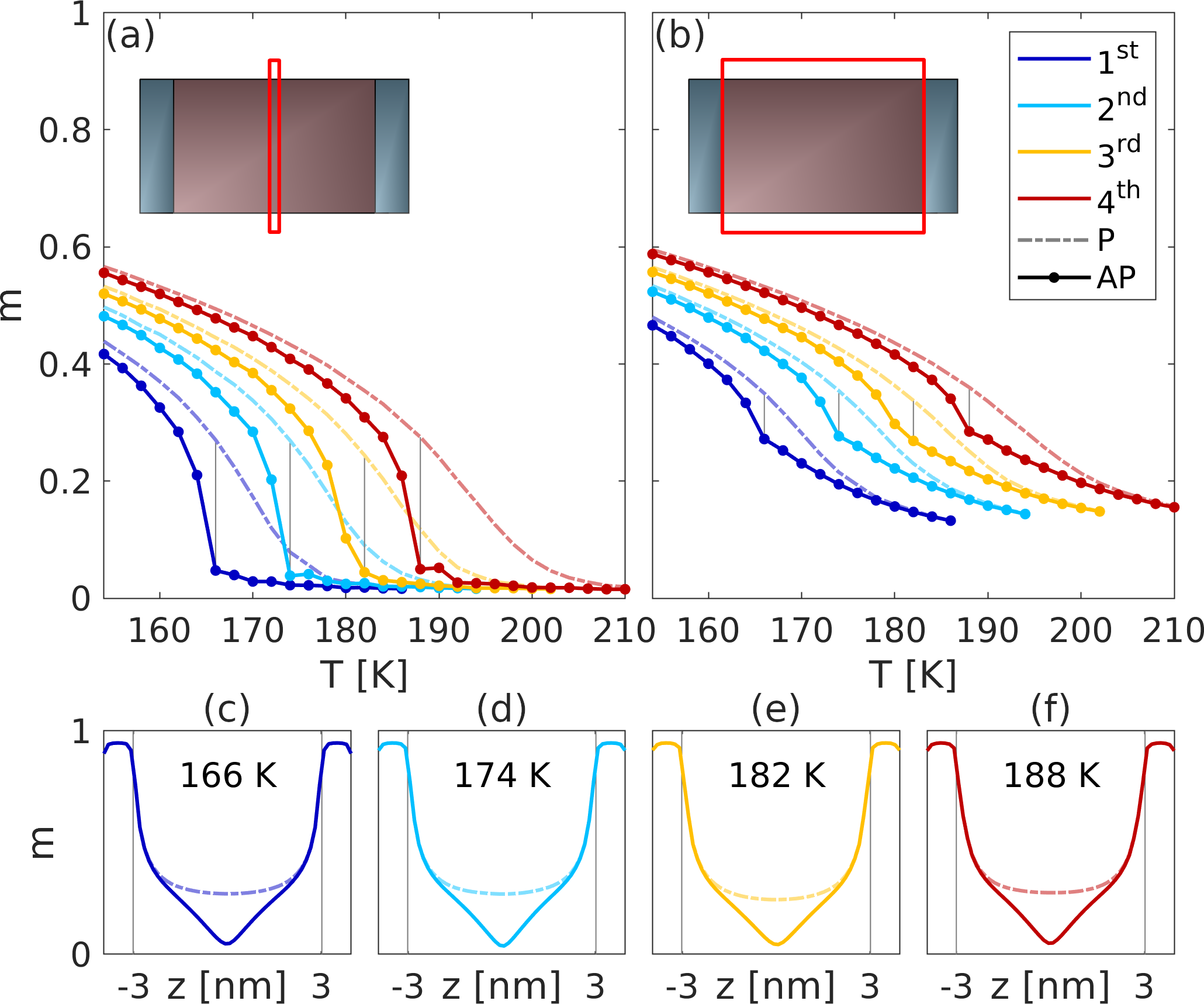}
\caption{\label{fig:m-v-T} (a) Magnetization at the center of the spacer at different temperatures in P and AP states, for four cases of spin-spin interaction range from 1st through 4th nearest neighbors. (b) Total magnetization of the whole spacer in P and AP states. (c-f) Spatial profiles of magnetization at temperatures of maximum magnetization difference P-vs-AP (marked by vertical lines in (a,b) panels and given as captions in (c-f) panels) for the four cases of exchange range [1st (c) through 4th n.n.\ (f)].}
\end{figure}

The system is relaxed at different temperatures from 0~K to well above the Curie temperature of the spacer (200 -- 214~K, depending on the interaction range), to find the maximum magnetization difference between the P and AP states. The temperature change is sequential with a step size of $\Delta T = 2$~K and at each temperature the system is equilibrated for 20~ps, followed by 50~ps of averaging. An example of the simulated magnetization profile for the AP state is illustrated in Fig.~\ref{fig:sample} and the spacer's magnetization versus temperature is shown in Fig.~\ref{fig:m-v-T}(a,b). This magnetization is not a projection as one would obtain in magnetometry, nor is it the magnitude of the total magnetization vector of the spacer. Instead it is the order parameter, $m$, extracted by taking the average normalized magnetization of all monolayers in the spacer.

The magnetization of the spacer in the P and AP states versus temperature follow one another quite closely until the AP magnetization abruptly drops approximately 10~K below the Curie transition in the P state. The longer exchange range can at this point only be seen to have one clear effect in that it increases the effective Curie temperature. This effect is only quantitative and do not change the overall behavior.

\begin{figure}[!t]
\centering
\includegraphics[scale=0.45]{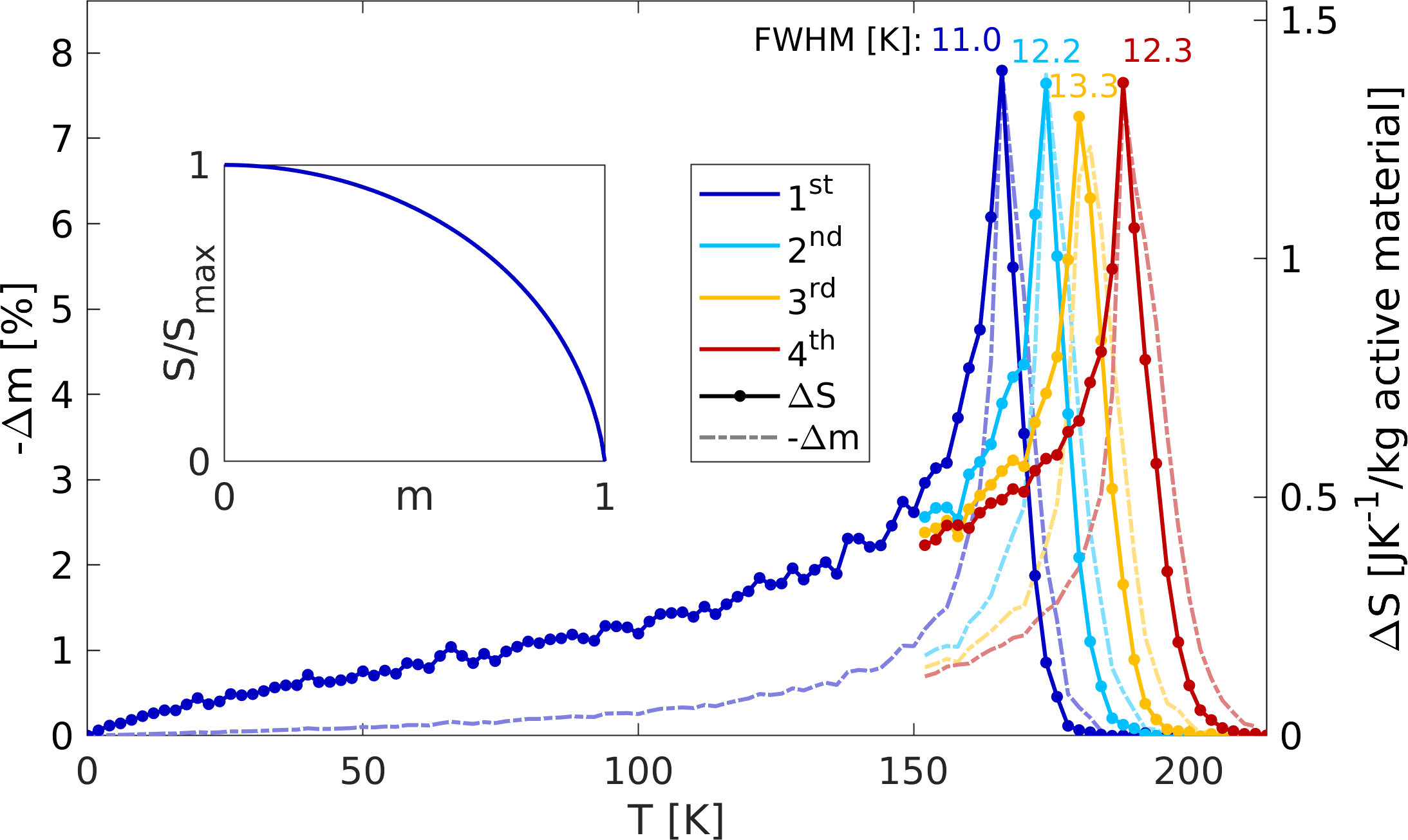}
\caption{\label{fig:dmAnddSvsT} Change in magnetization and entropy between P and AP states at different temperatures. Inset shows general relation between magnetization and entropy.}
\end{figure}

The magnetization difference on P to AP switching, $\Delta m$, is shown in Fig.~\ref{fig:dmAnddSvsT} as a function of temperature (as $-\Delta m$), for the same four cases of the spin-spin exchange range. In order to translate $\Delta m$ into a difference in entropy, the magnetization in each monolayer is considered before using the additive property of entropy, $\Delta S = \sum_i^N S(m_{\text{ap}}^i)-S(m_{\text{p}}^i)$.
For the entropy of a $(2J+1)$-state ferromagnet, where $J$ is the total angular momentum, a mean field model was presented by Tuszy\'{n}ski and Wierzbicki~\cite{Tuszy_ski_1991}, in which
\begin{equation}
    \begin{split}
        \frac{S_J(m)}{Nk_{\text{B}}} = \ln\frac{\sinh\big[(2J+1)x\big]}{\sinh(x)} + 2xJm,
    \end{split}
    \label{eqn:generalEntropy}
\end{equation}
where $x = g_{SLJ}\mu_{\text{B}}\nu/2$, proportional to the Lagrange coefficient $\nu$, is found by numerically solving the equation
\begin{equation}
    2Jm = \coth(x) - (2J+1)\coth\big[(2J+1)x\big].
\end{equation}
The above model is used with $S = 3/2$, $L = 0$, $J = S+L$, and $g_{SLJ} = 1.478$ for low temperature bulk iron~\cite{Koebler_2003}, and when converting from the entropy per-atom to per-mass of the active material, the spacer is considered to have 30\% magnetic (Fe) atoms, see Fig.~\ref{fig:dmAnddSvsT}. With $-\Delta m$ and $\Delta S$ plotted on the left and right axis, respectively, a difference in the shape of the curves is easily seen. $\Delta S$ is above $-\Delta m$ at low temperature and below it at high temperatures, since a given change in magnetization causes a larger entropy change near saturation compared to that close to zero magnetization. The corresponding $m$-vs-$S$ relation is shown in the inset. A consequence is that the peak in $\Delta S$ is at slightly lower temperature than the peak in $-\Delta m$.

Regarding the effect of the exchange range, it may appear that the peaks in Fig.~\ref{fig:dmAnddSvsT} become wider (quantified via FWHM and shown in respective color) as the $T_\text{C}$ increases, however the FWHM values are affected by this shift in temperature. After rescaling the temperature axes such that the four entropy peaks align, the widths become 11.0, 11.6, 12.1, and 10.9~K, with no clear trend present. Vertical rescalingof ($\Delta S$) does not affect the FWHM, but if a peak is truncated, then the FWHM is increased. This is why the FWHM value for the $3$rd n.n.\ case sticks out -- the 2~K increments of the temperature sweep `missed' the respective peak somewhat.

The effects of increasing the spin-spin interaction range up to 4th n.n.\ are rather intuitive and found above to be relatively weak, such that the qualitative behavior of the system, when it relates to MCE, does not change. In order to reduce the computational complexity of the atomistic modeling of the large phase space of the studied multilayer, in what follows the spin-spin interaction range is restricted to the 1st nearest neighbors, which nevertheless allows us to capture the main characteristics of interest. Long-range exchange, going far beyond the 4th n.n., and its relevance for MCE, calls for a somewhat less rigorous than spin-atomistics, but computationally much more compact, phenomenological numerical model, to be discussed elsewhere~\cite{Kulyk_2022}.

\subsection{Magnetization Profiles}
Different functional forms have been used to describe the profile of magnetization through the spacer in a F/f/F trilayer, including linear, hyperbolic sines and cosines, whose parameters are found by minimizing the free energy~\cite{Kravets_2014,Fraerman_2015,Kuznetsov_2021}. The assumptions and simplifications used when arriving at these expressions may be difficult to justify when the internal magnetization has not been (cannot be) measured and there are no direct MCE measurements to compare the final results to. Although the simulation model used here is idealized on a number of aspects, such as intra-layer homogeneity and no surface roughness, the key aspect of the spatial profile under thermal agitation is at the ultimate microscopic precision of the atomic mesh. The internal magnetization profiles in the exchange-proximity affected spacer we obtain numerically are used below to support more accurate functional $m(z)$ forms, which should prove useful for future analytical modeling of the magnetic properties in this and similar exchange-spring systems.

\begin{figure}[!t]
\centering
\includegraphics[scale=0.45]{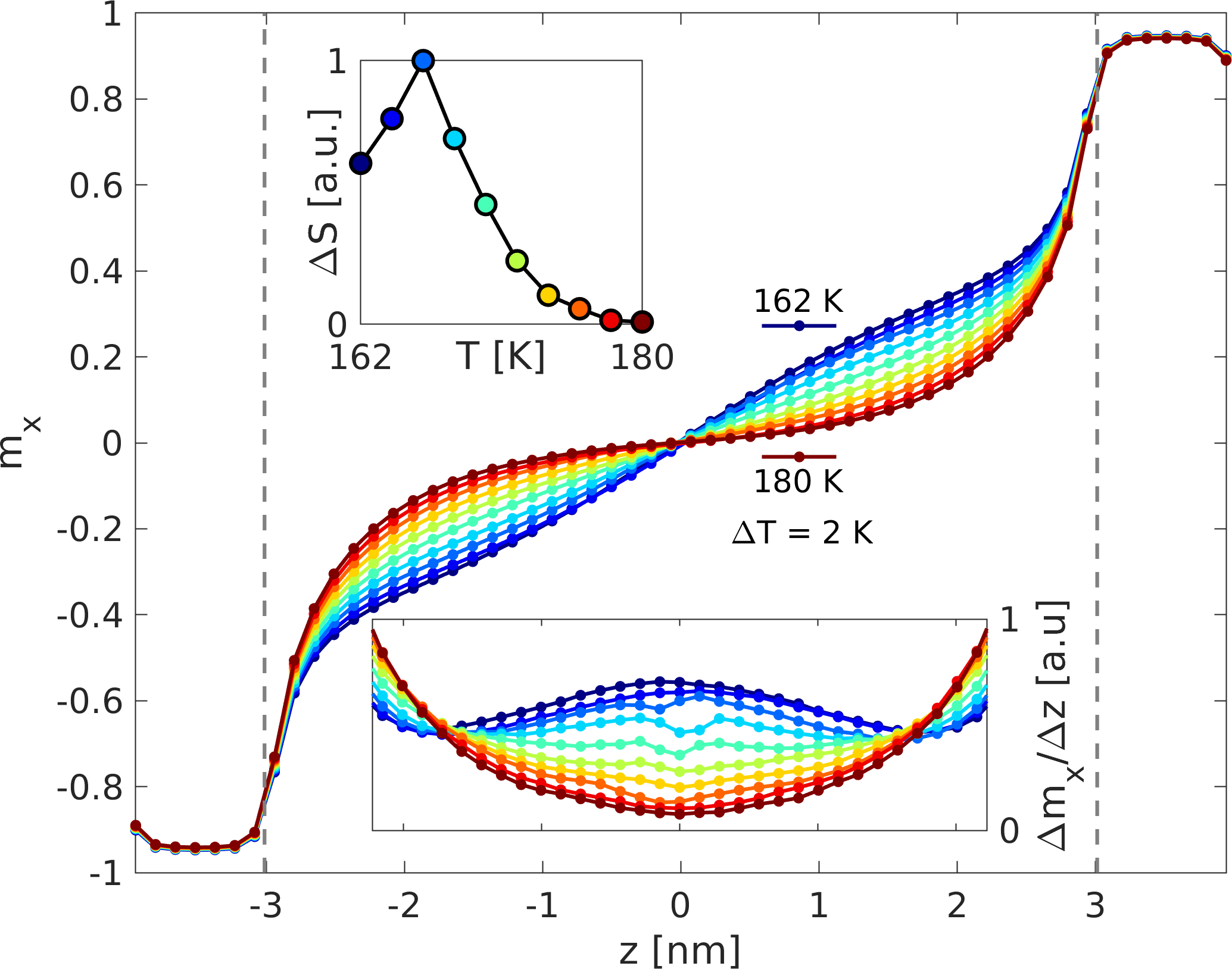}
\caption{\label{fig:profiles} Magnetization profiles in AP state of F/f/F system at temperatures near $T_{\text{C}}$ of the spacer. High-temperature profiles resembles those typically used in the literature, however maximum MCE is at $166$~K (top inset), where $m(z)$ is qualitatively different. Dashed lines mark F/f and f/F interfaces. (Bottom inset) Derivative of magnetization with respect to $z$.}
\end{figure}

The hyperbolic-cosine form employed in the literature~\cite{Kravets_2014,Kuznetsov_2021}, based on phenomenological considerations, fits our numerical results rather well for the P state of the system. For the AP state, however, our atomic spin profiles data differ qualitatively from the functional forms used previously, as illustrated in Fig.~\ref{fig:profiles}. Around the peak in magnetic disorder, $\Delta S_{\text{max}}$, and at lower temperatures, $|m(z)|$ has a cusp in the center of the spacer corresponding to a local maximum in the derivative, $m'(z)$, instead of a minimum in $m'(z)$ seen with hyperbolic-sine profiles. The form of $m(z)$ in our simulations approaches a hyperbolic sine as the temperature is increased, but not before $-\Delta m$ (MCE) has essentially vanished. This qualitative difference between our results and those obtained previously~\cite{Kravets_2014,Kuznetsov_2021} suggests that the first-order phenomenological assumption that the spacer is paramagnetic, while simplifying derivations, is not fully justified. A proper analytical derivation, treating the spacer as ferromagnetic (albeit weakly; even at $T>T_{\text{C}}$), will not be attempted here, but better fitting expressions for $m(z)$ are presented after the following paragraph, in which the spin-atomistically obtained profiles are made more intuitive.

\begin{figure}[!t]
\centering
\includegraphics[scale=0.45]{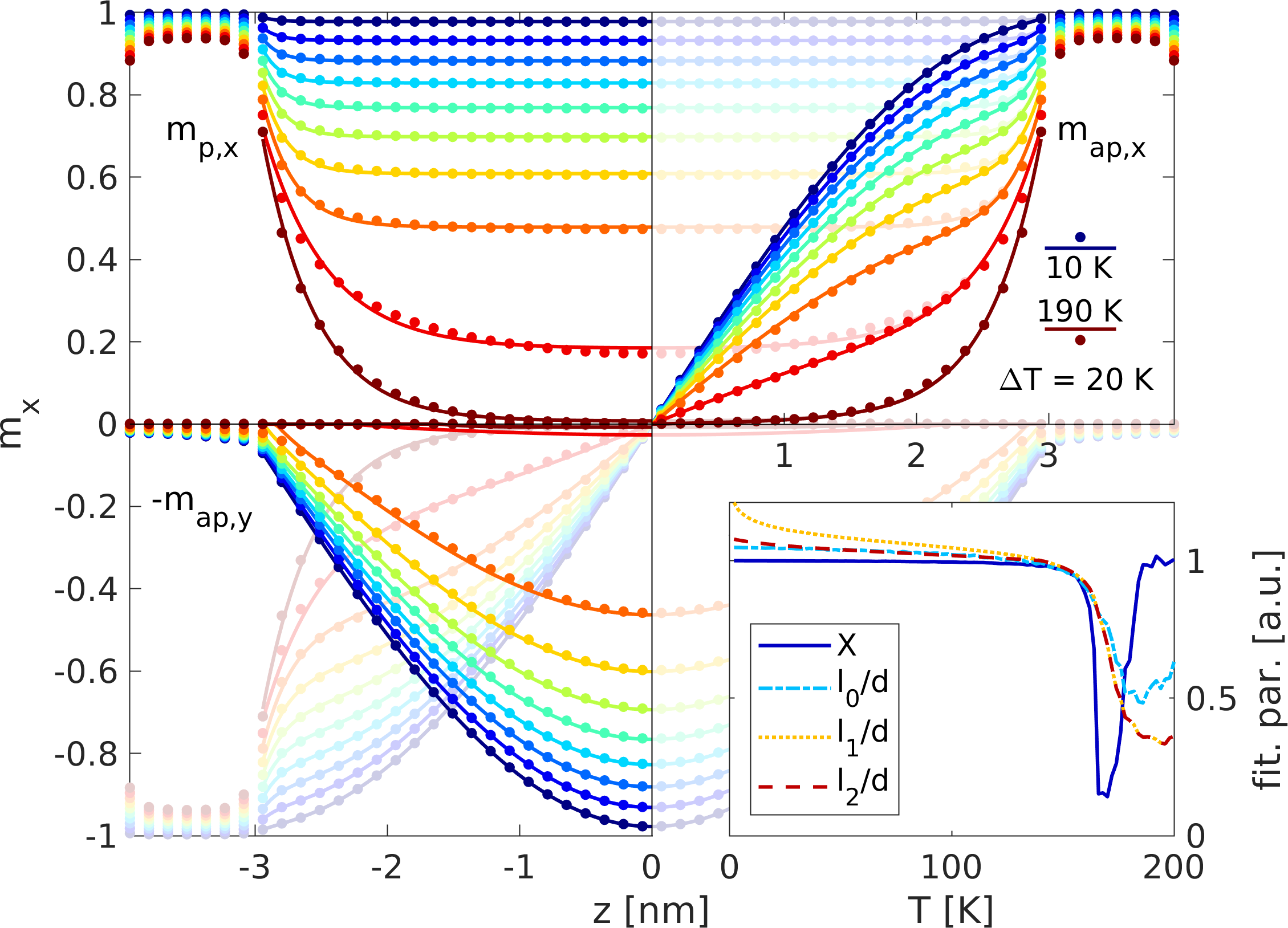}
\caption{\label{fig:projectionProfiles} Projections of magnetization profiles in P and AP states. Points are simulated data and curves are fits using Eq.~\eqref{eqn:fittingfunctions}. The symmetry of $m_{\text{p},x}$ and $m_{\text{ap},y}$ and antisymmetry of $m_{\text{ap},x}$ was used to display all data in this compact way. Inset shows the ratio $X=m_{\text{ap}}(z=0)/m_{\text{p}}(z=0)$, fitted domain wall length $l_0$, calculated $l_1$ (without boundary term), and $l_2$ (with boundary term).}
\end{figure}

The Bloch type domain wall present in the AP state at low temperatures is uniformly distributed across the spacer at 0~K. As the temperature increases and the magnetization magnitude in the center decreases due to thermal disorder, the domain wall shortens since it is less costly energetically in regions of lower magnetization. A shorter domain wall means a larger angle between the adjacent moments (monolayers' magnetizations), which in turn means a lower effective exchange field and, hence, additional reduction in the domain wall length. This eventually leads to a `collapse' of the domain wall and a peak in $-\Delta m$. There is still a domain wall near the center of the spacer after this peak temperature, but it is much less pronounced as its baring magnetization essentially vanishes under high thermal agitation. These characteristics of the simulated data suggest that the magnetization in the AP state at a given temperature is better described by a product of the hyperbolic cosine of the P-state $m(z)$ profile at that same temperature multiplied by the magnetization of the domain wall, whose strength and length decreases with increasing temperature:
\begin{equation}
\begin{split}
    m_{\text{p}}(z) &= m_{\text{p},x}(z) = \alpha_1\cosh\big(\alpha_2z\big) + \alpha_3,\\
	m_{\text{ap},x}(z) &= 
	\begin{cases} m_{\text{p}}(z),& |z| \geq l/2,\\
	                m_{\text{p}}(z)\sin\frac{\pi z}{l}, & |z| < l/2,
	\end{cases}\\
    m_{\text{ap},y}(z) &= 
    \begin{cases} 
    0,& |z| \geq l/2,\\
                    X m_{\text{p}}(z)\cos\frac{\pi z}{l}, & |z| < l/2,
    \end{cases}
\end{split}
\label{eqn:fittingfunctions}
\end{equation}
where $X = m_{\text{ap}}(0)/m_{\text{p}}(0)$. These expressions fit the simulated data well at all temperatures; see Fig.~\ref{fig:projectionProfiles}. First, parameters $\alpha_1$, $\alpha_2$, $\alpha_3$ are found by fitting the parallel state magnetization, and the resulting $m_{\text{p}}(z)$ is used for the fitting function, $m_{\text{ap},x}(l;z)$. Fitting the AP state magnetization gives the domain wall length, $l$, which in turn can be used for obtaining the perpendicular magnetization component, $m_{\text{ap},y}(X,l;z)$. Factor $X$ can either be fitted for or calculated from $m_{\text{ap}}(0)$ and $m_{\text{p}}(0)$, as was done here; see the inset to Fig.~\ref{fig:projectionProfiles}. The spacer has an even number of monolayers, therefore the magnetization at $z=0$ is found by linearly extrapolating $m(z)$ from the two closest monolayers on either side of $z=0$, and then taking the average of these extrapolations.

For these expressions to be more than fitting functions and useful for indirect MCE measurements (often done using magnetometry), the fitting parameters would ideally be expressed via physical parameters that describe the system and the constituent materials. This has already been done for the parameters of $m_{\text{p}}(z)$~\cite{Kravets_2014,Kuznetsov_2021} in such a way that they can be estimated from hysteresis loops, although it was the case of a paramagnetic spacer that was considered in these articles. The problem of finding $\Delta m$ is then reduced to finding $m_{\text{ap}}(z)$, i.e.\ the parameters $X$ and $l$. For fixed values of $X$ the domain wall length can be found by minimizing the exchange energy,
\begin{equation}
    F_{\text{ex}} = \int_0^{d/2}\left(\frac{\text{d}\bm{m}}{\text{d}z}\right)^2\text{d}z,
    \label{eqn:exchangeEnergy}
\end{equation}
where no exchange parameter is included since it does not affect $\text{d}F/\text{d}l = 0$. Zeeman energy need not be considered as no external field is applied within the spacer. Shape anisotropy is not a factor either since the functional form $\bm{m}(z)$ already restricts the magnetic moments to lie in the plane of the spacer. Specifics of exchange at the interfaces are excluded as they are not important, due to the sharp growth of the hyperbolic cosine function near the interfaces, which helps restrict the domain wall to the spacer.

Splitting integral~\eqref{eqn:exchangeEnergy} in two, for the intervals $(0,l/2)$ and $(l/2,d/2)$, and taking the derivative with respect to $l$ leads to, for a fixed $X\in(0,1]$,
\begin{equation}
    \begin{split}
        &\frac{l}{2}m_{\text{p}}^2\left(\frac{l}{2}\right) = 2\int\limits_0^{l/2}m_{\text{p}}^2(z)\left[\sin^2\Big(\frac{\pi z}{l}\Big) + X^2\cos^2\Big(\frac{\pi z}{l}\Big)\right]\text{d}z\\
        &\quad+(1-X^2)\int\limits_0^{l/2}z\sin\Big(2\frac{\pi z}{l}\Big)\left[\frac{l}{\pi}m_{\text{p}}^{'2}(z) + \frac{\pi}{l}m_{\text{p}}^2(z)\right]\text{d}z.
    \end{split}
    \label{eqn:generalSolution}
\end{equation}
This equation can be solved numerically for $l$, given $X$, and the results are displayed in the inset to Fig.~\ref{fig:projectionProfiles}, denoted $l_1$. The $X = m_{\text{ap}}(0)/m_{\text{p}}(0)$ ratio was obtained from the simulated spin profiles (non-trivial to obtain analytically). To achieve better agreement at low temperatures, a boundary term of the form $c_b(\bm{m}_{\text{f}} - \bm{m}_{\text{F}})^2|_{z=d/2}$ was added to the free energy~\cite{Fraerman_2015}, which, with a large enough coefficient $c_b$, can represent the high cost of the domain wall extending into the magnetically hard outer F-layers. The improved results are displayed in the inset to Fig.~\ref{fig:projectionProfiles}, denoted $l_2$.

The above detailed analysis of the thermodynamics of the Bloch-type domain wall in the AP state is motivated by the key role it plays in the F/f/F system's MCE, which peaks when this wall collapses in length and magnetization near the spacer's $T_\text{C}$.

\begin{figure*}[!t]
\centering
\includegraphics[scale=0.45]{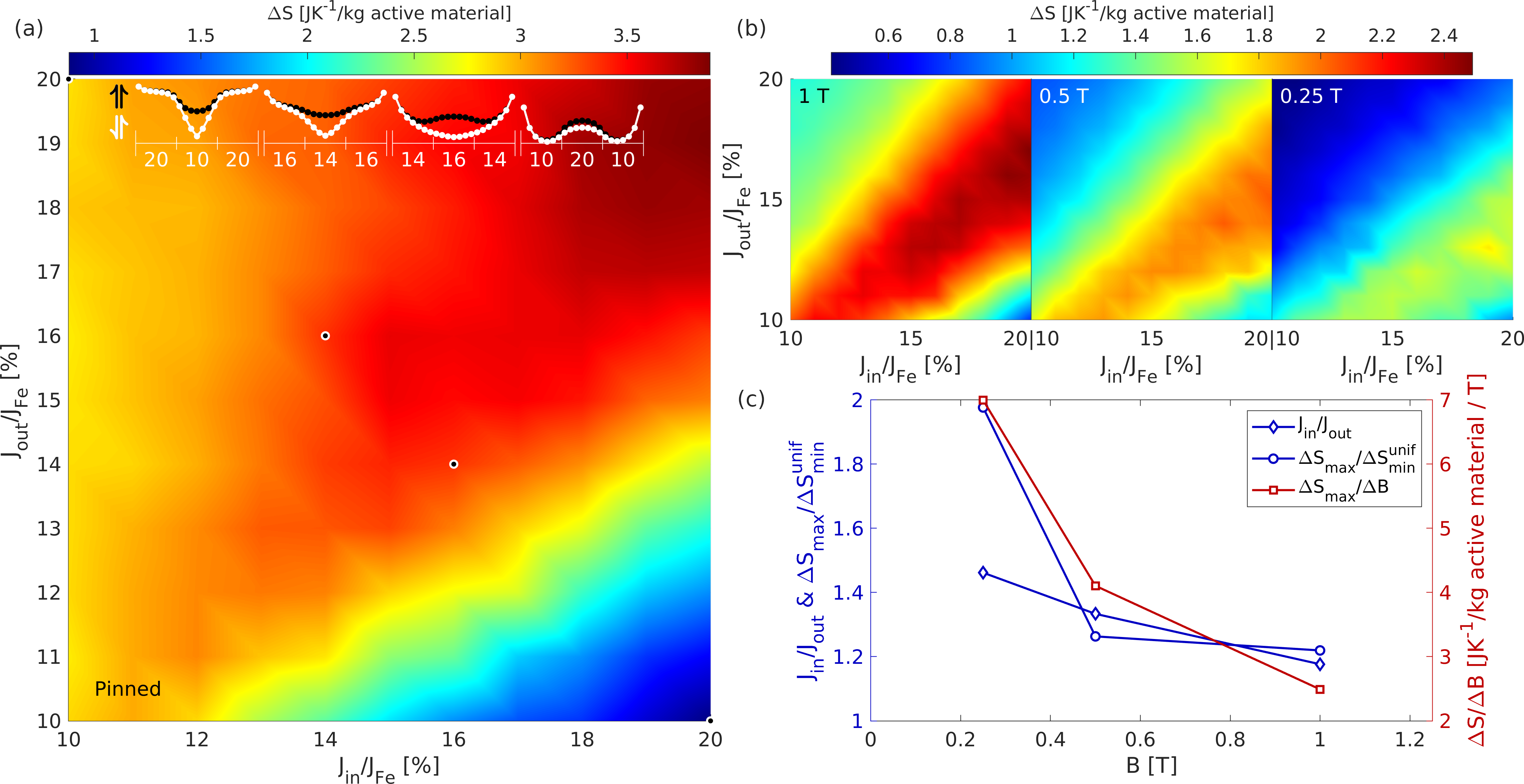}
\caption{\label{fig:entropyMap_composite} (a) Map over $\Delta S$ between P and AP states in F/f/F trilayers with gradient spacer exchange. Axes show exchange of outer and inner sections of the spacer. Profiles at maximum $\Delta S$ are shown at the top, for four selected spacers labeled by their corresponding exchange values and marked on the map (dots along 20-20 diagonal). Magnetization of the outer ferromagnetic layers is fixed with strong anisotropy. (b) $\Delta S$ maps with anisotropy in the free layer replaced by external fields (labels). (c) Ratio $J_{\text{in}}/J_{\text{out}}$ corresponding to maximum $\Delta S$ versus external field; ratio $\Delta S_{\text{max}}/\Delta S_{\text{min}}^{\text{unif}}$, quantifying the improvement of using non-uniform spacers; and $\Delta S_{\text{max}}$ per unit field.}
\end{figure*}

\subsection{MCE Optimization}
When studying the profiles of magnetization and the difference between the P and AP states, the question arises of whether it would be possible to increase $\Delta m$ by, for example, lifting up the parallel state magnetization in the center of the spacer. The first attempt to improve the system could therefore be to use a spacer with a spatially varying Curie temperature, realized experimentally by varying the concentration of the magnetic component in the spacer alloy, similar to how it was done by A.\ F.\  Kravets \textit{et al.}~\cite{Kravets_2012}.

The spacer is divided into three equal sections, and the exchange in the inner section $J_{\text{in}}$ is allowed to vary relative to the exchange in the outer sections $J_{\text{out}}$. The two parameters are swept from 10\% $J_{\text{Fe}}$ to 20\% $J_{\text{Fe}}$ in steps of one percentage point, such that the ratio $J_{\text{in}}/J_{\text{out}}$ varies from 0.5 to 2. This amount of simulations requires a large amount of time and computational resources, even when restricting the exchange to nearest neighbors, and so the spacer thickness was reduced from 6~nm to 3~nm. At each point $(J_{\text{in}},J_{\text{out}})$ a temperature sweep is performed in order to find the maximum entropy change, which represents the spacer in the resulting map $\Delta S(J_{\text{in}},J_{\text{out}})$. For the calculation of the entropy at any point $(J_{\text{in}},J_{\text{out}})$ one must have not just the order parameter $m = m(z)$, but also the saturation magnetization, which should depend on the exchange values since they represent concentration of magnetic material. The relations between the concentration, saturation magnetization, and Curie temperature have been well studied for alloys such as Fe$_x$Cr$_{1-x}$~\cite{Burke_1983_1,Burke_1983_2,Burke_1983_3,Ravi_Kumar_2015,Ravi_Kumar_2018}, and could be used to find $M_s(J_{\text{in}},J_{\text{out}};z)$. However, this would make the results less general and, for consistency and simplicity, a linear approximation is used, with 15\% exchange corresponding to 30\% iron atoms, as for the system above with the uniform 6~nm spacer, and 100\% exchange corresponding to 100\% iron atoms. The saturation magnetization, for the calculation of entropy, was then set to be proportional to the magnetic concentration, just as the atomic moments were in the simulations.

The resulting $\Delta S(J_{\text{in}},J_{\text{out}})$ can be seen in Fig.~\ref{fig:entropyMap_composite}(a), and the largest values lie just below the line $J_{\text{in}}/J_{\text{out}} = 1$, i.e.\ the best spacer is nearly uniform. However, this is with the outer ferromagnets pinned by anisotropy fields of 55~T, meaning these $\Delta S$ values are the largest achievable for each spacer and the external field required to saturate $\Delta S$ to these values may vary with the exchange values. To investigate this, the 55~T anisotropy in the free layer is replaced by a much weaker external field. Just as the anisotropy was localized to the free layer, the external field is too.

With a field of 1~T, there is a clear change in the optimal ratio $J_{\text{in}}/J_{\text{out}}$ in favor of the original idea of lifting up the magnetization in the center of the spacer, i.e.\ $J_{\text{in}}/J_{\text{out}} > 1$; see Fig.~\ref{fig:entropyMap_composite}(b). The largest $\Delta S$ are comparable to those obtained with the high anisotropy, but the entropy per unit field has increased by at least an order of magnitude. The change in the $\Delta S$ surface suggests that 1~T is below the saturation field for $(J_{\text{in}},J_{\text{out}})$ that are above the ridge of maximum $\Delta S$, but still above the saturation field for points below this ridge. Therefore, the field is further reduced by 2 and 4 times, to 0.5~T and 0.25~T; Fig.~\ref{fig:entropyMap_composite}(b). At 0.5~T, the largest $\Delta S$ per unit field is observed at $(20\%,15\%)$ and is 26\% greater than the minimum for a uniform spacer, while for 0.25~T these numbers are $(19\%,13\%)$ and 98\% -- a significant enhancement of a factor of two. It is important to note that this two-fold enhancement is achieved with less Fe in the gradient spacer (31.6\% Fe atoms for the case of 0.25~T reversing field) compared to the uniform spacer case (minimum found at 34.1\% Fe atoms). 

Units of entropy per field are not meant to imply that $\Delta S$ scales linearly with the field, but if a smaller field can reorient the ferromagnetic layers, then more re-orientations can be performed per unit time, and $(\Delta S/\Delta t)/(\Delta B/\Delta t) = \Delta S/\Delta B$.

For large ratios $J_{\text{in}}/J_{\text{out}}$, the maximum $\Delta m$ occurs when the magnetization is near zero in the outer sections of the spacer, and the difference is concentrated in the central section. Most of the domain wall will then be located in the outer sections, while the center is more uniform. This is problematic as it means that an external field, which on the experiment is not localized to the free layer, could have a significant effect on the central magnetization in the AP state, and thereby counter-act the MCE. Using a thicker free layer would reduce this problem and increase the MCE per unit field, but it would also reduce the fraction of active material (spacer vs total volume). Hence, the best way to further improve the system would be to look for more efficient exchange (magnetic dilution) profiles $J(z)$. For any ratio $J_{\text{in}}/J_{\text{out}}$, there should be an optimal shape of the profile $J(z)$ with $J(0) = J_{\text{in}}$ and $J(|z| = d/2) = J_{\text{out}}$.

\section{Conclusions}
Highly parallelized computation applied to atomistic spin dynamics, in contrast to algorithmically serial Monte-Carlo methods, makes it possible to quickly find the spatially resolved magnetization (microscopic spin distribution) in realistic and computationally rather large nanosystems. The systems' entropy can be calculated in a straightforward way, directly from the spin distribution using a mean field model for ferromagnets with identical spins.
Extending the exchange interaction range from the 1st to the 4th nearest neighbors results in quantitative changes of the simulated properties, with the overall qualitative picture intact. Much longer range spin-spin interactions, suggested by some of the experiments for the investigated metallic F/f/F system, where the conduction band is polarized and can mediate exchange over several nanometers, in addition to the more localized lattice-exchange, are computationally too heavy for spin-atomistic numerics and call for specialized phenomenological modeling.

The simulated magnetization profiles in the parallel and antiparallel states are studied in detail. While the parallel state is well described by hyperbolic cosines, frequently used in the literature, it is found that in order to describe the antiparallel state the domain wall present at low temperatures must be taken into consideration. Even at high temperatures, after the domain wall appears to have collapsed, it is effectively present, and a combination of a hyperbolic cosine and a domain-wall function that varies in strength and length, fits the antiparallel state magnetization at all temperatures. By using the hyperbolic cosine for the parallel state at the same temperature, the problem of finding the antiparallel state magnetization reduces to finding the domain wall length and the ratio of the parallel and antiparallel state magnetizations at the center of the spacer. If the latter is known, then the domain wall length can be found by numerically solving an equation free from the material parameters, which is equivalent to minimizing the free energy.

Finally, the effects of using a nonuniform spacer are studied, with the exchange, or concentration of magnetic material, allowed to vary through the spacer. With the spacer divided into three equal sections it was found that the entropy change per unit field was significantly improved when the inner section was given a larger exchange than the outer sections. Further improvements should be possible with more sophisticated exchange (concentration) profiles.

\begin{acknowledgments}
Support from the Swedish Research Council (VR 2018-03526) and the Olle Engkvist Foundation (project 2020-2022) are gratefully acknowledged. The computations were enabled by resources provided by the Swedish National Infrastructure for Computing (SNIC), partially funded by the Swedish Research Council through grant agreement no.\ 2018-05973. We thank Lars Viklund (HPC2N) and Tor Kjellsson (PDC) at SNIC for their assistance with installation, use and modification of software, which was made possible through application support provided by SNIC.
\end{acknowledgments}

\bibliography{MCE_directExchange_references}

\end{document}